# *In Vivo* Quantitative Analysis of Anterior Chamber White Blood Cell Mixture Composition Using Spectroscopic Optical Coherence Tomography

**RUOBING QIAN,**[1,*] **RYAN P. MCNABB,**[2] **KEVIN C. ZHOU,**[1] **HAZEM M. MOUSA,**[2] **DANIEL R. SABAN,**[2] **VICTOR L. PEREZ,**[2] **ANTHONY N. KUO,**[2,1] **AND JOSEPH A. IZATT**[1,2]

[1]*Department of Biomedical Engineering, Duke University, Durham, NC 27708, USA*
[2]*Department of Ophthalmology, Duke University Medical Center, NC 27710, USA*
*\*ruobing.qian@duke.edu*

**Abstract:** Anterior uveitis is the most common form of intraocular inflammation, and one of its main signs is the presence of white blood cells (WBCs) in the anterior chamber (AC). Clinically, the true composition of cells can currently only be obtained using AC paracentesis, an invasive procedure to obtain AC fluid requiring needle insertion into the AC. We previously developed a spectroscopic optical coherence tomography (SOCT) analysis method to differentiate between populations of RBCs and subtypes of WBCs, including granulocytes, lymphocytes and monocytes, both *in vitro* and in ACs of excised porcine eyes. We have shown that different types of WBCs have distinct characteristic size distributions, extracted from the backscattered reflectance spectrum of individual cells using Mie theory. Here, we further develop our method to estimate the composition of blood cell mixtures, both *in vitro* and *in vivo*. To do so, we estimate the size distribution of unknown cell mixtures by fitting the distribution observed using SOCT with a weighted combination of reference size distributions of each WBC type calculated using kernel density estimation. We validate the accuracy of our estimation in an *in vitro* study, by comparing our results for a given WBC sample mixture with the cellular concentrations measured by a hemocytometer and SOCT images before mixing. We also conducted a small *in vivo* quantitative cell mixture validation pilot study which demonstrates congruence between our method and AC paracentesis in two patients with uveitis. The SOCT based method appears promising to provide quantitative diagnostic information of cellular responses in the ACs of patients with uveitis.

## 1. Introduction

Anterior uveitis, inflammation of the iris and the ciliary body, is the most common form of intraocular inflammation [1], with an estimated 14–17 uveitis cases per 100,000 persons confirmed each year in the USA [1]. In some cases, uveitis can lead to serious complications such as cataract, glaucoma and cystoid macular edema [2], which accounts for 10% of legal blindness in the USA [3]. Despite its importance as a vision threatening disease, the causes of many anterior uveitis cases are unknown [4]. One of the main signs of anterior uveitis is the presence of white blood cells (WBCs) in the anterior chamber (AC). The clinical standard evaluation is grading of these cells by uveitis specialists on a scale of 0 to 4+ using slit-lamp microscope examination, as defined by the Standardization of Uveitis Nomenclature (SUN) [5]. One limitation of slit-lamp evaluation is that subjective qualitative assessment of the condition makes it difficult to judge the longitudinal course of the cellular response. Another major limitation is the lack of differentiation between subtypes of WBCs, including granulocytes, lymphocytes and monocytes. Many human and animal studies have shown that different etiologies of uveitis have different cell composition patterns. For instance, predominantly granulocytic responses may suggest acute and autoimmune inflammations such as HLA-B27-asscoiated acute uveitis [6, 7] , while a predominant response of mononuclear WBCs, mainly lymphocytes and monocytes, may suggest sarcoidosis [8], Vogt–Koyanagi–Harada [9], or melanin-protein-induced uveitis [10] as examples. Therefore, the composition of AC cells can potentially provide additional diagnostic information regarding the underlying cause of uveitis and hence guide management. Clinically, this information regarding the true composition of cells can currently only be obtained using AC paracentesis which requires invasive needle insertion into the anterior chamber to withdraw fluid with potential complications [11, 12].

Optical coherence tomography (OCT) allows noninvasive, high-resolution cross-sectional imaging of the AC [13], and can be a useful adjunct to provide quantitative diagnostic information regarding cellular responses associated with uveitis. Several groups have previously developed technologies to localize and count cells appearing in anterior segment OCT (AS-OCT) images in human or animal models [14-18]. However, none of these technologies were able to provide quantitative information regarding the composition of cellular mixtures. Although different types of WBCs have distinct cell sizes or nucleus shapes and sizes, the differences cannot be distinguished directly with contemporary clinical AS-OCT systems because the axial resolution of these systems, typically 5-10μm, is comparable to the size of WBCs. Therefore, individual blood cells appear as indistinguishable hyper-reflective spots in conventional AS-OCT images [19]. Rose-Nussbaumer et al. have previously demonstrated the possibility to differentiate subtypes of WBCs using the reflectance intensity in OCT images [19]. However, the cell reflectance in the OCT images can possibly be affected by many external factors, such as illumination laser power, optical alignment, and cornea opacities.

To address this shortcoming, we have previously reported a spectroscopic OCT (SOCT) analysis approach to differentiate subtypes of WBCs, including granulocytes, lymphocytes and monocytes, both *in vitro* and within *ex vivo* porcine eyes [20]. Our approach involved tracking and extracting the backscatter spectrum of isolated single blood cells in repeated OCT B-scans or volumes and correlating the spectral features of single cells to their characteristic sizes using Mie theory fitting. We demonstrated that all three subtypes of WBCs had statistically distinct size distributions. In this work, we further develop this method to estimate the composition of unknown AC cellular mixtures. We have validated the accuracy of our method in both *in vitro* experiments and a preliminary *in vivo* clinical pilot study.

## 2. Methods

*2.1* In vitro *cell mixture study*

### 2.1.1 OCT system design

*In vitro* imaging was performed utilizing a custom 100-kHz swept-source OCT (SSOCT) system, centered at 1052 nm with a bandwidth of 108 nm (Axsun Technologies; Billerca, MA) with custom imaging optics and OCT engine, as detailed in the previous study [20]. The system had measured axial and lateral resolutions of ~8.8 μm and ~20 μm respectively, and a maximum axial imaging range of 7.4 mm. The measured sensitivity and -6-dB fall off of the system was 103 dB and 4.39 mm with 2.2 mW illumination power on the sample.

### 2.1.2 *In vitro* cell mixture preparation

Cryopreserved granulocytes and lymphocytes from healthy donors (ZenBio Inc.; Research Triangle Park, NC) were obtained; these types were chosen as they are the two predominant WBC types. The detailed procedure we used to obtain cell suspensions in glass cuvettes for *in vitro* imaging was detailed in our previous study [20]. To summarize, cryopreserved WBC samples (~15 million cells per viral) were rapidly thawed in 37$^{\circ}$C water bath and re-suspended in 15 mL PBS solutions containing 0.1% bovine serum albumin after the cryoprotectant was removed.

We then created sample mixtures of WBCs consisting of granulocytes and lymphocytes. Before mixing the two cell-type samples, two independent approaches were used to determine the true concentrations of the samples. The concentration of each sample was first determined by adding 25μL of cell suspension into a hemocytometer (Thermo Fisher Scientific; Waltham, MA), and counting the total number of cells using a 10x microscope objective. As a second, complementary method, OCT was used to separately determine the concentration of each granulocyte or lymphocyte sample. Single B-scans with 500 A-scans/B-scan and the same field of view (FOV) of 2 mm were acquired at 20 different locations for each sample, and the previously developed cell localization algorithm [20] was applied to count the total number of cells in each scanned dataset. Known volumes of the two component cell samples were then well mixed with different ratios to create three sample mixtures with target granulocyte-lymphocyte ratio of 75%:25%, 60%:40% and 40%:60%, and ~1mL of each cell mixture (~1 million cells in total) was placed in a glass cuvette for SOCT imaging.

### 2.1.3 Data acquisition

Due to the weak backscattering signal from single blood cells, repeated measurements from the same cell were required in order to obtain the spectrum of an individual cell with sufficient SNR for signal processing. We previously developed single cell localization and tracking algorithms for both repeated B-scan and volumetric OCT imaging [20]. In this study, only the repeated B-scan mode was used given the shortened time constraints that would be required for later *in vivo*/clinical imaging experiments. We placed the sample on a manual translation stage (Thorlabs,inc; Newton, NJ) aligned with the sample such that the observed dominant cell movement was in the same direction as the fast scanning axis, and 200 repeated B-scans with 500 A-scans/B-scan and a field-of-view (FOV) of 2 mm were acquired at 6 different locations of each cell mixture.

### 2.1.4 Single cell tracking and spectroscopic analysis

The methods to track and extract the backscattering spectrum of isolated single blood cells and correlate the spectral features of single cells to their characteristic sizes were detailed in the previous paper [20]. To summarize, we located individual cells in consecutive B-scans by intensity thresholding and labeling connected components. We used the A-scan at the center of every labelled cell for the following spectroscopic analysis. Between any two adjacent B-scans, if a cell located in the second B-scan was within ~80 μm around a cell in the previous B-scan, these two cell images were recorded as arising from the same cell. When the peak cell intensity fell below a defined threshold value, the cell was recorded as having exited the B-scan. Due to the low backscattering signal from single cells, only cells staying in same FOV for at least 15 repeated B-scans were used for the following spectroscopic analysis. We extracted the backscatter spectrum of each cell from its center A-scan using the short-time Fourier transform (STFT). The STFT here uses a Hamming window with a spectral bandwidth of 23.7 cm$^{-1}$, corresponding to an axial resolution of ~340 μm for each window, and the window was shifted across the acquired interferogram a total of 100 times with a step size of 9.2 cm$^{-1}$. We then normalized the STFT backscatter spectrum by dividing by the known laser power spectrum. After that, we obtained the background spectrum of each cell from the adjacent A-scans without cells at the same depth and performed noise subtraction. Finally, we used Mie theory to correlate the backscatter spectra of single cells to their characteristic sizes (which are likely the size of the nucleus [21]).

### 2.1.5 Estimation of cell mixture composition

We have previously shown that all three subtypes of WBCs, granulocytes, lymphocytes and monocytes, have distinct characteristic size distributions [20]. In this study, we further verified our results on more cell samples from different cryopreserved healthy donor cells and used size distributions acquired from these samples as the reference size distributions for each subtype of WBCs (results in section 3.1). We calculated the probability density functions (PDFs) of size distributions using kernel density estimation [22],

$$PDF(x) = \frac{1}{nh}\sum_{i=1}^{n} K(\frac{x-x(i)}{h}) \quad (1)$$

with a Gaussian kernel function $k(u) = \frac{1}{\sqrt{2\pi}}exp(-\frac{1}{2}u^2)$. Here $x(i)$ is the characteristic size of cell $i$ in the distribution, $n$ is the total number of the cells in the distribution, and $h$, the bandwidth of the kernel function, was chosen to be 0.01μm. We then estimated the cell composition by minimizing the mean square error (MSE) between the PDF of our cell mixture sample ($PDF_{mix}$), and a weighted combination of the two component reference PDFs, $PDF_{gran}$ and $PDF_{lympho}$,

$$MSE = \frac{1}{s_{max}-s_{min}}\int_{s_{min}}^{s_{max}}\bigl(PDF_{mix}(s) - (P_{gran}*PDF_{gran}(s) + P_{lympho}*PDF_{lympho}(s))\bigr)^2 ds \quad (2)$$

where s is the characteristic size, $s_{min}$ and $s_{max}$ are the minimum and maximum characteristic sizes in the distribution, $P_{gran}$ and $P_{lympho}$ are the concentration percentages of granulocytes and lymphocytes in the cell mixture, and $P_{gran} + P_{lympho}$ =100%. One-sample t-tests were then performed between our estimation results and the two reference methods results to validate the accuracy of our method.

*2.2 In vitro cell mixture simulation study*

Beside the *in vitro* cell mixture study, we also performed a simulation study to predict the accuracy of our cell mixture composition quantification method. Here, we first generated a simulated characterize size dataset of a total of $N_{gran}$ granulocytes or $N_{lympho}$ lymphocytes. To achieve that, we first calculated the cumulative distribution function (CDFs) of each cell type by integration of the reference probability density functions (PDF) over size. We then generated N random numbers from the uniform distribution on the interval (0,1), and found the corresponding characterize size of each random simulated cell through the inverse CDF. To match the cell population statistics of our *in vitro* sample, characteristic sizes of a total of 1,000,000 cells including 1,000,000* $T_{gran}$ granulocytes and 1,000,000* $T_{lympho}$ lymphocytes were simulated, where $T_{gran}$ and $T_{lympho}$ are the true concentration percentages of granulocytes and lymphocytes in the cell mixture and $T_{gran} + T_{lympho}$ =100%. Here, $T_{gran}$=20%, 40%, 60% and 80% were chosen to simulate from granulocyte-dominated to lymphocyte-dominated cell mixture. Then, we randomly sampled *M* cells from this simulated cell mixture population, calculated the probability density functions (PDF) of the sampled sample size distribution, and estimated the cell composition $P_{gran}$ and $P_{lympho}$ using the method just described in 2.1.5. We repeated this procedure 500 times to obtain the means and standard deviations of $P_{gran}$ and $P_{lympho}$.

*2.3 In vivo pilot study of uveitis patients*

2.3.1 OCT system for *in vivo* pilot study

To facilitate *in vivo* imaging, a custom high-speed SSOCT system was built for the *in vivo* pilot study (Fig. 1) at the Duke Eye Center. The system employed a 200-kHz swept-wavelength laser (Axsun Technologies; Billerca, MA), centered at 1050 nm with a sweep range of 100 nm. The OCT system utilized a transmissive Mach-Zehnder interferometer topology and the interferometric signal was detected by a balanced photoreceiver (Insight

Photonic Solutions, Inc.; Lafayette, CO) with 400-MHz electronic bandwidth and digitized at 800 MS/s (AlazarTech Inc.; Pointe-Claire, QC, Canada). The axial imaging range of the system was 7.4 mm in air. The lateral resolution of the system was designed for 12.8 μm FWHM and 840 μm depth of focus in air. To image throughout the anterior chamber, the system incorporated a focusing module in the sample arm (Fig. 1) to allow for the adjustment of the focal plane axially. The module consisted of two achromatic lenses configured in a 1:1 4F telescope configuration where the first lens was located on a motorized translation stage (National Aperture, Inc.; Salem, NH). The functionality of this module is detailed in the following section 2.3.2.

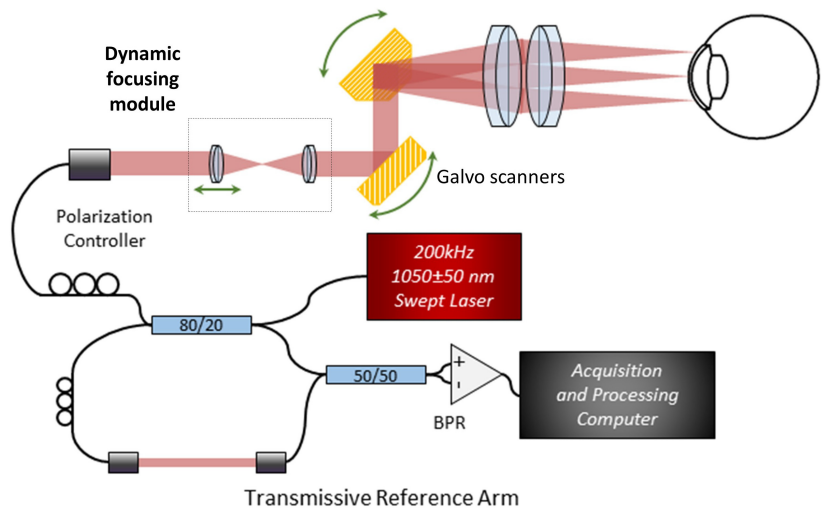

Fig.1 Schematic diagram of the custom 200kHz SS-OCT system with focusing module to adjust the focal plane axially. BPR: balanced photoreceiver.

### 2.3.2 *In vivo* imaging

The following pilot study was approved by the Duke University Medical Center Institutional Review Board, and adhered to the Health Insurance Portability and Accountability Act and all tenets of the Declaration of Helsinki. Patients with clinically active anterior uveitis of at least 2+ on slit lamp microscopy performed by a uveitis specialist on the day of imaging were enrolled after obtaining written informed consent. The optical power incident on the cornea at 1050 nm was measured at or below 4.1 mW before each imaging session, which is under the maximum permissible exposure determined by American National Standards Institute (ANSI) safety standards (ANSI Z80.36-2016).

The primary difference between all the prior *in vitro*/*ex vivo* experiments and the *in vivo* imaging was the effect of patient motion during each measurement, as our approach requires multiple OCT B-scans or volumes to obtain the averaged cell spectrum with sufficient SNR for spectroscopic analysis. Therefore, it was important to adjust the imaging scan parameters for *in vivo* imaging at the beginning of the study, with the goal of keeping each single cell within the FOV in consecutive scans while at the same time maximizing the FOV to capture more cells, so that the total imaging time was minimized. A total of 7 uveitis patients were imaged during this optimization stage. We successfully imaged and tracked >100 cells in 3 of 7 patients. Too few cells were localized or successfully tracked in the other four patients.

We quickly identified that our prior scan protocol based on repeated volumes [20] with the utilized acquisition rate was not ideal for *in vivo* imaging, as >10 repeated OCT volumes combined with a reasonable number of A-scans/B-scan and B-scans/volume required at least several seconds of imaging, and motion artifacts significantly compromised the cell tracking algorithm. Through imaging initial subjects, we also observed that cells were not uniformly distributed in each subject's AC, so it was important to identify regions with increased cell density at the beginning of the imaging session. Therefore, we designed the following imaging protocol for our study.

Each imaging session started with wide-field volumetric imaging to cover the FOV of nearly the whole AC to study the blood cell distribution in the AC. Once the regions with dense distributions of cells were identified on the real-time OCT volumetric images, we translated the slit-lamp base and the focusing module in the sample arm to adjust the axial focal plane and lateral position of imaging to these regions. Then, a scanning protocol of 800 A-scans/B-scan, 15 repeated B-scans at 10 different positions along the slow scanning axis (150 B-scans in total), and a FOV of 2 mm by 1 mm, was used. For each subject, we repeated this protocol for 4-5 times to ensure that enough cells were available for cell composition estimation.

After imaging parameter optimization, two uveitis subjects for which paracentesis of the AC fluid (a relatively rare procedure) was indicated were enrolled and imaged in the quantitative cell mixture validation pilot study. The true compositions of the AC cells obtained from paracentesis were compared to our estimation results.

### 2.3.3 *In vivo* data processing

The methods to track and extract the spectrum of single cells was similar to the *in vitro* study described above and the previous paper [20], except the minimum number of B-scans (averages) for analysis was reduced from 15 to 6 to ensure that cells stayed within the FOV. Similar to the method in 2.1.5, the spectroscopic OCT based cell composition of each clinical subject was estimated by minimizing the MSE between the PDF of our *in vivo* size distribution ($PDF_{sub}$) and a weighted combination of all three component reference PDFs, $PDF_{gran}$(granulocyte), $PDF_{lympho}$(lymphocyte) and $PDF_{mono}$(monocyte).

$$MSE = \frac{1}{s_{max}-s_{min}} \int_{s_{min}}^{s_{max}} \big(PDF_{sub}(s) - (P_{gran} * PDF_{gran}(s) + P_{lympho} * PDF_{lympho}(s) + \\ + P_{mono} * PDF_{mono}(s))\big)^2 ds \tag{3}$$

where s is the characteristic size, $s_{min}$ and $s_{max}$ are the minimum and maximum characteristic sizes in the distribution, $P_{gran}$, $P_{lympho}$ and $P_{mono}$ are the concentration percentages of granulocytes, lymphocytes and monocytes in the cell mixture, and $P_{gran} + P_{lympho} + P_{mono}$ =100%. Here, $P_{lympho}+P_{mono}$ also corresponds to the concentration percentage of mononuclear WBCs.

### 2.3.4 AC fluid cell analysis

The AC fluid of the first enrolled uveitis subject was sent to a clinical pathology lab and stained using a modified Papanicolaou stain. The slide was then read by a clinical research scientist. The slide was scanned systematically using a 40x microscope objective. The cells were manually counted to identify the number of granulocytes, lymphocytes and monocytes. Nucleus shape was the main basis for differentiating the types of cells. Cells were counted twice to confirm numbers were accurate and representative.

The AC fluid of the second subject was sent to a research lab, where the sample was simultaneously immunostained for CD45, CD11b, CD11c, CD3, CD4, CD14, CD15, CD16, HLA-DR, and Live/Dead. Data were acquired on a BD Fortessa SORP and analyzed to enumerate live lymphocytes, monocytes and granulocytes using FlowJo software (TreeStar Inc., Ashland, OR).

## 3. Results

### 3.1 Reference size distributions of subtypes of WBCs

In our previous paper [20], we demonstrated that granulocytes, lymphocytes and monocytes have distinct characteristic size distributions. Here we imaged additional cell samples from different cryopreserved healthy donor cells to further validate our observations, and we used size distributions acquired from these cells as the reference size distributions for subsequent cell mixture studies. Histogram plots of characteristic sizes obtained from approximately two thousand cells of each cell type are shown in Fig. 2, along with their means and standard deviations. The means and standard deviations here are similar to our previously reported values [20] (granulocyte: 4.1 μm vs. 4.1 μm; lymphocyte: 5.6 μm vs. 5.7 μm; monocyte: 6.6 μm vs. 6.8 μm). These three reference size distributions were used to estimate the mixture cell composition in the following sections.

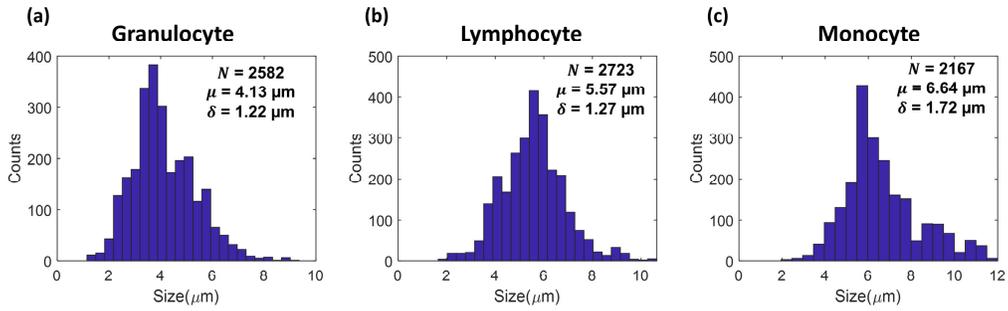

Fig. 2 Histogram plots of characteristic cell sizes of (a) granulocyte (b) lymphocyte and (c) monocyte extracted from the best Mie spectra fit with mean and standard deviation. N: total number of cells, μ, δ : mean and standard deviation of the characteristic size distribution

### 3.2 In vitro cell mixture simulation results

With the obtained reference size distributions of granulocytes and lymphocytes, we calculated the cumulative distribution function (CDFs) through integration the reference probability density functions (PDFs) over size, as shown in Fig. 3(a). In Fig. 3(b), we showed the simulated characteristic size distribution of a sample of 1,000,000 cells consisting of 600,000 (60%) granulocytes and 400,000 (40%) lymphocytes. Here, we studied the accuracy of our method as a function of number of imaged/sampled cells, M, and true cell composition, $T_{gran}$ and $T_{lympho}$. To accomplish this, we varied the number of imaged cells, M, from 100 to 2000 at four different true cell composition levels, $T_{gran}$ = 20%, 40%, 60% and 80%, and plotted the means and standard deviations of $P_{gran}$ (Fig. 3(c-d)). As expected, the standard deviations of $P_{gran}$ decrease as M increases, since both sampling error and fitting error are reduced with an increased number of imaged cells. More importantly, the standard deviations of $P_{gran}$ are very similar at different levels of $T_{gran}$(Fig.3(d)), which indicates that the accuracy of our method appears independent of the actual composition of the cell mixture.

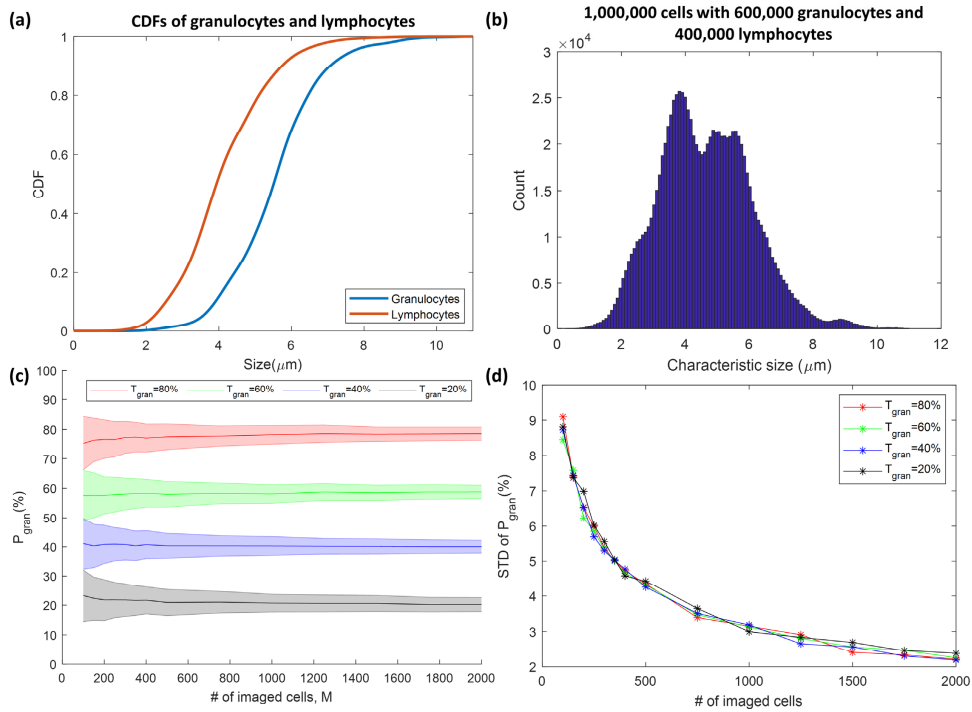

Fig.3 (a) Reference cumulative distribution function (CDF) of granulocytes and lymphocytes. (b) Histogram plot of simulated characteristic cell sizes of 1,000,000 cells with 600,000 (60%) granulocytes and 400,000 (40%) lymphocytes. (c) Estimated composition percentages of granulocytes, $P_{gran}$, as a function of the number of imaged cells at four different levels of $T_{gran}$. Banded curves represent the mean ± standard deviation (SD). (d) Plot of SD of $P_{gran}$ as a function of imaged cells at four different levels of $T_{gran}$, illustrating that the cell mixture prediction accuracy appears independent of the mixture proportion.

### 3.3 In vitro *cell mixture results*

Repeated B-scans were taken at 6 different locations for each cell mixture sample. Individual cells were localized and tracked, and the corresponding characteristic size of each identified cell based on the best Mie spectra fit was extracted from the averaged spectrum. A representative B-scan from cell mixture sample #1 with color-coded characteristic sizes of each of the individual cells is shown in Fig. 4(a). Cells without color coding correspond to cells that did not remain in the FOV for at least 15 repeated B-scans. A histogram plot of the characteristic sizes of all the identified cells (N=338) at this location is shown in Fig. 4(b), along with their mean and standard deviation. The extracted backscattering spectrum, best-fit Mie spectrum, and extracted characteristic sizes of two individual cells in the representative B-scan are shown in Fig. 4(c-d). The composition of the cell mixture at this FOV was found to be 63.2% lymphocytes and 36.8% granulocytes by fitting the characteristic size PDF with a weighted combination of two reference PDFs (Fig. 4(e)). Similarly, the representative characteristic size PDFs and the best fits of the weighted combination of reference PDFs of the other two mixture samples are shown in Fig.4 (f-g), along with the corresponding composition estimation results. The cell composition estimation results at all 6 locations using our spectroscopic OCT method are shown in Table 1, and the averaged lymphocyte and granulocyte proportions across all 6 locations of three mixture samples were found to be 62.1% & 37.9%, 40.0% & 60.0%, and 24.5% & 75.5% with standard deviations of 5.7%, 3.5% and 3.0%. We then compared the estimation results with the results obtained from the component sample counts using the hemocytometer and OCT prior to mixing. Overall, the estimated results using our spectroscopic OCT method agreed well with the results from the two reference methods, with percentage deviations of < 1.5% for the OCT reference and <2.5 % for the hemocytometer reference. One-sample t-tests were performed between our estimation results and the two reference method results, and no statistically significant difference was found between the estimation results of our spectroscopic OCT method and the two reference methods among all three cell mixtures ($p$ = 0.58, 0.84, 0.91 for the OCT reference and $p$ = 0.98, 0.19, 0.19 for the hemocytometer reference). Moreover, the measured cell composition from OCT images before mixing agreed with the results from the hemocytometer counts with a deviation of less than 2.5%, indicating that our single cell localization algorithm was capable of identifying most of the cells.

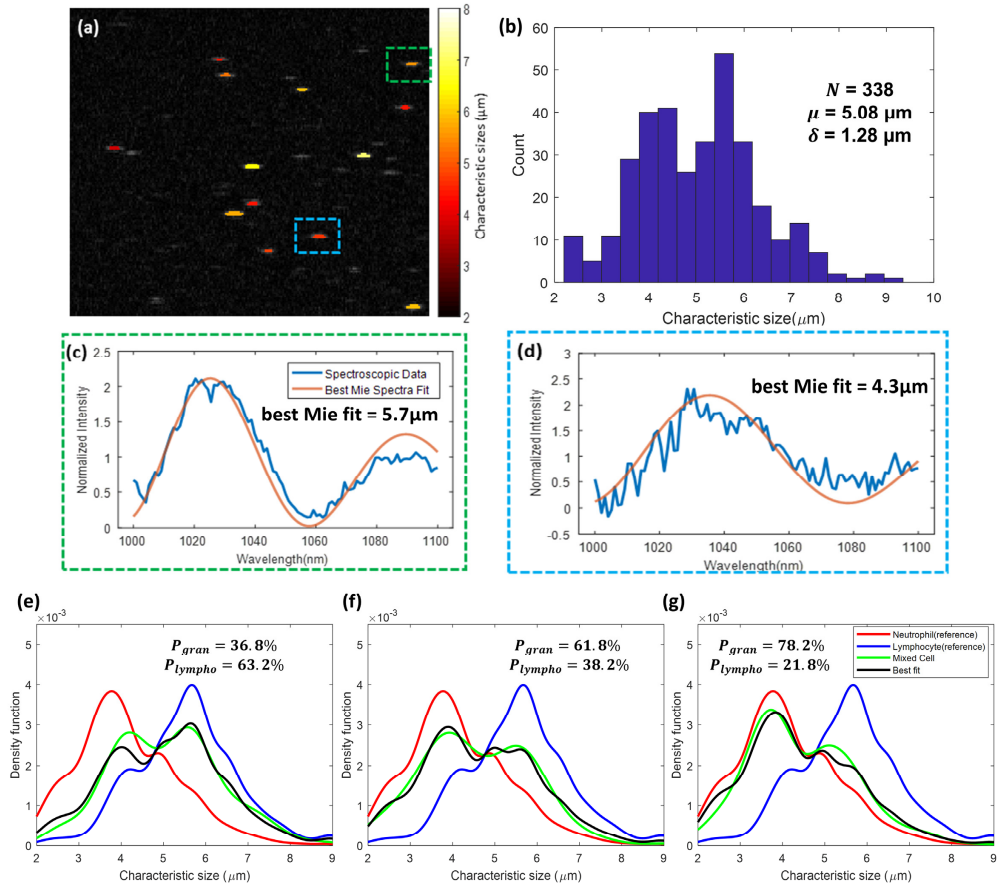

Fig.4 Results of the *in vitro* cell mixture study. (a) Overlay of characteristic sizes of individual cells with an OCT B-scan image of a cell mixture sample. (b) Histogram plot of characteristic cell sizes extracted from the best Mie spectra fit with mean and standard deviation. (c-d) Representative spectra of two single cells and their best Mie spectra fit. (e-g) Best fits (black) of representative characteristic size PDFs of three mixture samples (green) with a weighted combination of reference size PDFs of granulocytes (red) and lymphocytes (blue).

We also compared the experimental cell composition results here with the simulation results in section 3.2. The standard deviations of the experimental results of three cell mixtures, 5.7%, 3.5% and 3.4%, are similar to the standard deviation of simulation results when a total of 350 cells were imaged, ~5%.

**Table.1 Summary of the *in vitro* cell mixture composition estimation results**

| Target lymphocyte-granulocyte ratio | | \multicolumn{3}{c}{Cell Mixture #1} 60% :40% | | | \multicolumn{3}{c}{Cell Mixture #2} 40%:60% | | | \multicolumn{3}{c}{Cell Mixture #3} 25%:75% | | |
|---|---|---|---|---|---|---|---|---|---|---|---|
| Method | FOV | N | $P_{lympho}$ (%) | $P_{gran}$ (%) | N | $P_{lympho}$ (%) | $P_{gran}$ (%) | N | $P_{lympho}$ (%) | $P_{gran}$ (%) |
| Spectroscopic OCT | 1 | 338 | 63.2 | 36.8 | 318 | 38.2 | 61.8 | 418 | 21.8 | 78.2 |
| | 2 | 369 | 56.7 | 43.3 | 351 | 45.5 | 54.5 | 384 | 30.3 | 69.7 |
| | 3 | 325 | 68.0 | 32.0 | 352 | 39.9 | 60.1 | 353 | 26.3 | 73.7 |
| | 4 | 303 | 60.4 | 39.6 | 388 | 37.4 | 63.6 | 346 | 21.2 | 78.8 |
| | 5 | 346 | 55.3 | 44.7 | 319 | 36.2 | 63.8 | 347 | 23.4 | 76.6 |
| | 6 | 362 | 68.9 | 31.1 | 309 | 42.8 | 57.2 | 339 | 24.9 | 75.1 |
| | Mean (SD) | | **62.1 (5.7)** | **37.9 (5.7)** | | **40.0 (3.5)** | **60.0 (3.5)** | | **24.6 (3.4)** | **75.4 (3.4)** |
| Intensity counts by OCT before mixing (20 averaged) | | | **60.7** | **39.3** | | **39.7** | **60.3** | | **24.8** | **75.2** |
| | | \multicolumn{3}{c}{***p* = 0.58**} | | | \multicolumn{3}{c}{***p* = 0.84**} | | | \multicolumn{3}{c}{***p* = 0.91**} | | |
| Hemocytometer (3 averaged) | | | **62.1** | **37.9** | | **42.2** | **57.8** | | **26.7** | **73.3** |
| | | \multicolumn{3}{c}{***p* = 0.98**} | | | \multicolumn{3}{c}{***p* = 0.19**} | | | \multicolumn{3}{c}{***p* = 0.19**} | | |

### 3.3 In vivo pilot study results

For the first enrolled subject, a representative OCT B-scan with individual cells localized is shown in Fig. 5(a) and (b). We applied our spectroscopic analysis method to extract the spectrum and calculate the characteristic size of individual cells. Representative spectra of two localized cells and their best Mie spectra fits are shown in Fig. 5(c) and (d). A total of 234 cells were successfully tracked and measured in this subject, and the histogram plot of characteristic cell sizes with mean and standard deviation is shown in Fig. 5(e). We then estimated the cell composition by minimizing the MSE between the measured PDF of the size distribution and a weighted combination of three reference PDFs (granulocytes, lymphocytes and monocytes) (Fig. 5(f)). For this subject, we estimated the cell composition to be 66.0% granulocytes and 34.0% mononuclear WBCs (lymphocytes: 25.2%; monocytes: 8.8%), a granulocyte vs. mononuclear WBC ratio of 1.94:1. This ratio was confirmed by the AC fluid analysis result (granulocyte vs. mononuclear WBCs =1.5:1 - 2:1).

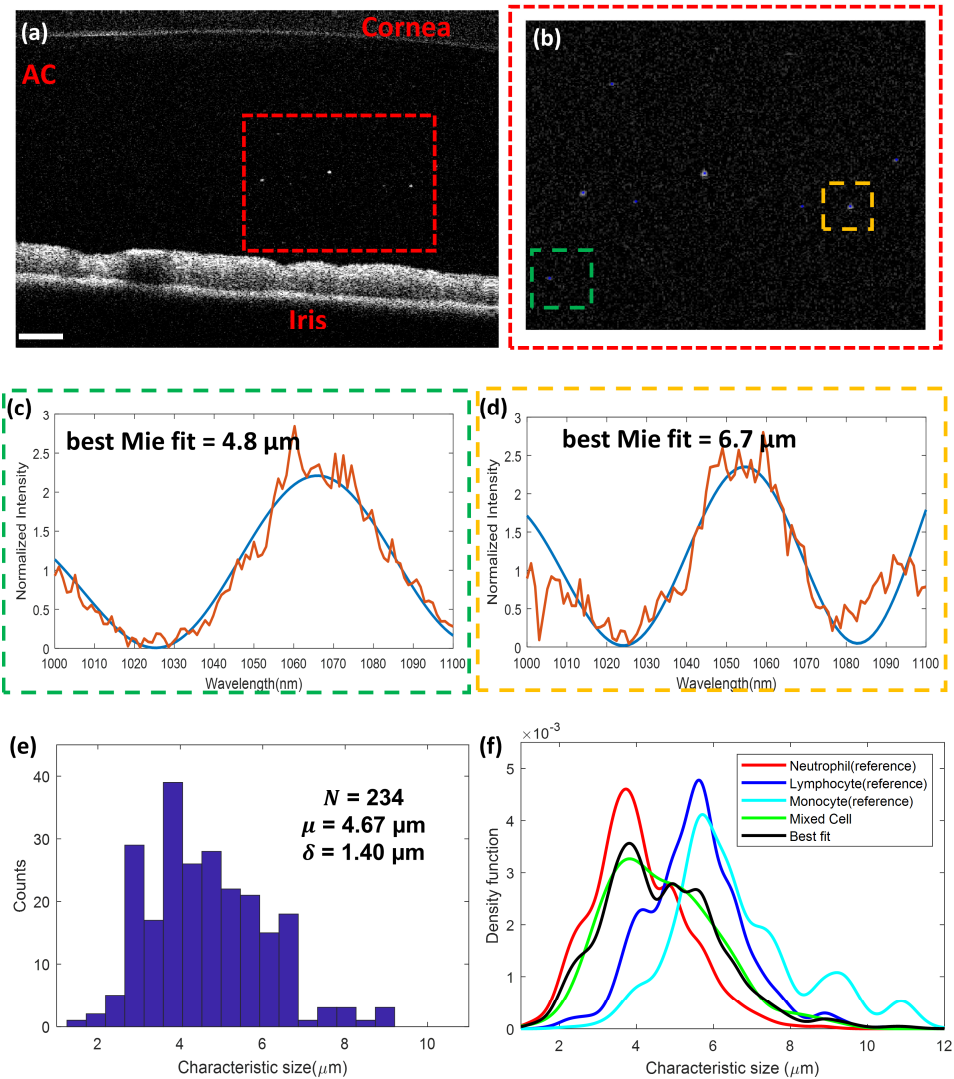

Fig.5 Results of cell composition analysis of the uveitis patient #1 in the *in vivo* pilot study. (a) A representative OCT B-scan of anterior chamber, and (b) a zoomed-in view with individual blood cells localized. (c-d) Two representative single cell spectra with their best Mie spectra fit. (e) Histogram plot of characteristic cell sizes extracted from the best Mie spectra fit with means and standard deviations. (f) Best fit (black) of measured characteristic size PDF with a weighted combination of reference size PDFs of granulocytes (red), lymphocytes (blue) and monocytes (cyan).

For the second enrolled uveitis subject, a total of 458 cells were successfully identified and measured in this subject, and the histogram plot of characteristic cell sizes with mean and standard deviation is shown in Fig.6 (a). Similarly, we estimated the cell composition by fitting the measured size distribution with a weighted combination of three reference PDFs (Fig. 6(b)). The cell composition was estimated to be 15.0% granulocytes and 85.0% mononuclear WBCs (lymphocytes: 25.3%; monocytes: 59.7%), while the AC paracentesis has the following results: monocytes: 67%, lymphocytes: 22%, granulocytes: 1% and unclassified lymphoid population or unidentified cells: 10%. Although our estimated composition granulocyte is higher than the true composition (15.0% vs. 1%), the estimated compositions of lymphocytes and monocytes are close to the true compositions (59.7% vs. 67% and 25.3% vs.22%) with an error of only 7.3% and 3.3%.

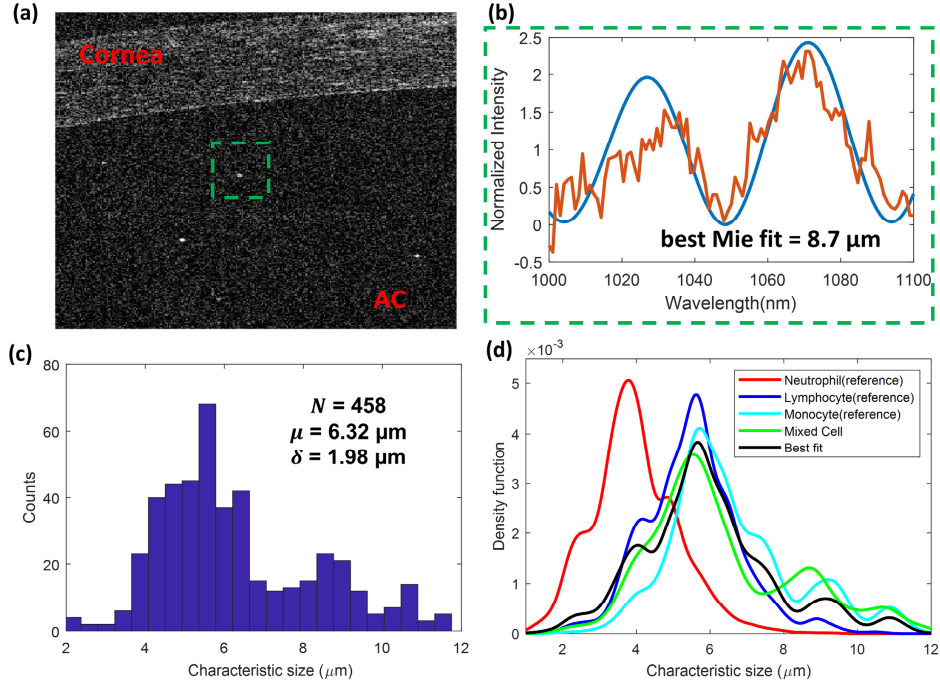

Fig.6 Results of cell composition analysis of the uveitis patient #2 in the *in vivo* pilot study. (a) A representative OCT B-scan of anterior chamber. (b) A representative single cell spectrum with their best Mie spectra fit. (c) Histogram plot of characteristic cell sizes extracted from the best Mie spectra fit with means and standard deviations. (b) Best fit (black) of measured characteristic size PDF with a weighted combination of reference size PDFs of granulocytes (red), lymphocytes (blue) and monocytes (cyan).

## 4. Discussion

We have previously shown that all three subtypes of WBCs, granulocytes, lymphocytes and monocytes, have distinct characteristic size distributions. The size distribution of each type of cell was found to overlap with the size distributions of other types of cells, which means it is challenging to differentiate cell type at the single-cell level. However, as demonstrated in this study, we still can estimate the true composition of mixtures of these three cell types, by fitting the measured size distribution of cell mixture samples to a linear combination of reference size distributions of cells.

A major challenge of our technology for *in vivo* human imaging are artifacts caused by patient motion. In the *in vitro* mixture cell studies, our cell tracking algorithm was able to localize and track the majority of cells in the FOV, as individual cells had only one dominant direction of motion (direction of gravity), and we aligned the system's B-scan direction accordingly to mitigate this motion effect. However, significant eye motion artifacts were present in *in vivo* imaging which compromised the performance of our tracking algorithm. The eye motion effects were more obvious when multiple repeated volumes with a large FOV were taken, which had an acquisition time of longer than several seconds. To reduce the effect of motion artifacts, only repeated B-scan mode with a small FOV (~2 mm) was used in our quantitative cell mixture validation pilot study. However, shortening the imaging time by reducing the FOV limits the number of cells which can be identified and analyzed from each acquisition. The total number of identified cells determines the estimation accuracy of our cell composition, as a larger number of identified cells lead to a more accurate representation of the characteristic size distribution of the cell population. As a result, in our pilot quantitative cell mixture validation study, only uveitis patients with a slit-lamp grading of >2+ were enrolled, and multiple acquisitions of repeated B-scans at different locations of the anterior chamber were performed in order to acquire a large enough number of tracked cells.

Multiple approaches can potentially improve the number of tracked cells in each measurement and thus improve the composition estimation accuracy. First, as we started to explore in our *in vivo* pilot study, optimized OCT imaging parameters (e.g. #of A-scans/B-scan, # of repeated B-scans, FOV) can minimize the motion effects between repeated B-scans and maximize the number of tracked cells in a single measurement. More quantitative studies need to be performed to find the optimal imaging protocol for in vivo imaging. Secondly, during the post-processing, parameters like intensity thresholding value were sometimes adjusted based on the illumination power, subject or sample alignment. These parameters along with the cell tracking algorithm between sequential frames can also be further optimized, with the goal to maximize the number of identified cells appeared in the AC. Moreover, real-time eye tracking and motion compensation techniques could also significantly reduce the motion effects, such as pupil tracking [23], a robotically-aligned OCT scanner [24] or SLO-based eye tracking [25].

In our study, only WBCs from healthy donors were used to build up the reference size distributions of subtypes of WBCs in the *in vitro* studies. However, activated lymphocytes caused by the inflammatory responses such as anterior uveitis may increase in size [26], and the reference size distributions of lymphocytes obtained from the *in vitro* studies may not be a very accurate representation of lymphocytes appearing in the AC, which may lead to errors in cell composition estimation. In future studies, the size distribution of lymphocytes obtained from the AC paracentesis could be directly studied using our method and compared to the size distribution of lymphocytes obtained from the healthy donors.

## 5. Conclusion

We have previously shown that spectroscopic OCT analysis of single blood cells can differentiate subtypes of WBCs, including granulocytes, lymphocytes and monocytes. In this paper, we further developed our method to estimate the composition of AC blood cell mixture. The characteristic size distribution of the cell mixture sample was obtained by identifying, tracking and extracting the backscattering spectrum of single blood cells in the repeat OCT B-scans, and correlating the spectral features of single cells to their characteristic sizes using Mie theory.  The probability density function of the size distribution was calculated using kernel density estimation, and fit with a weighted combination of reference size distributions of each WBC type to estimate the composition. We validated the accuracy of our method in an *in vitro* study, by comparing our estimations with two reference methods, hemocytometer and OCT images before mixing. Finally, we conducted an *in vivo* pilot study and demonstrated that our method can provide quantitative information of AC blood cell composition. The estimation results using our method demonstrated good agreement with the results from AC paracentesis. This work can potentially enable noninvasive quantitative diagnosis of cellular responses in the uveitis patients in the clinic.

## 6. Funding, acknowledgments, and disclosures

*6.1 Funding*

We acknowledge funding from USAMRAA grant W81XWH-16-1-0498, NIH Core Grant EY005722, and an Unrestricted Award from Research to Prevent Blindness, Inc. to Duke University. KCZ was supported by the National Science Foundation (DGF-1106401)

*6.2 Acknowledgements*

We thank Rose Mathew for the support of the hemocytometer experiment.

*6.3 Disclosures*

RPM: Leica Microsystems (P). ANK: Leica Microsystems (P). JAI: Leica Microsystems (P, R), Carl Zeiss Meditec (P, R).